\documentclass[sigconf, usenames, dvipsnames]{acmart}

\usepackage{dblfloatfix} 
\usepackage{booktabs} 
\usepackage{arydshln}
\usepackage{subfig}
\AtBeginDocument{%
  \providecommand\BibTeX{{%
    \normalfont B\kern-0.5em{\scshape i\kern-0.25em b}\kern-0.8em\TeX}}}

\copyrightyear{2025}
\acmYear{2025}
\setcopyright{rightsretained}
\acmConference[CHI EA '25]{Extended Abstracts of the CHI Conference on Human Factors in Computing Systems}{April 26-May 1, 2025}{Yokohama, Japan}
\acmBooktitle{Extended Abstracts of the CHI Conference on Human Factors in Computing Systems (CHI EA '25), April 26-May 1, 2025, Yokohama, Japan}\acmDOI{10.1145/3706599.3706741}
\acmISBN{979-8-4007-1395-8/25/04}

\begin{document}
\title{Everyday AR through AI-in-the-Loop}

\author{Ryo Suzuki}
\affiliation{%
  \institution{University of Colorado Boulder}
  \city{Boulder}
  \country{USA}}
\email{ryo.suzuki@colorado.edu}

\author{Mar Gonzalez-Franco}
\affiliation{%
  \institution{Google}
  \city{Seattle}
  \country{USA}}
\email{margon@google.com}

\author{Misha Sra}
\affiliation{%
  \institution{UC Santa Barbara}
  \city{Santa Barbara}
  \country{USA}}
\email{sra@cs.ucsb.edu}

\author{David Lindlbauer}
\affiliation{%
  \institution{Carnegie Mellon University}
  \city{Pittsburgh}
  \country{USA}}
\email{davidlindlbauer@cmu.edu}

\renewcommand{\shortauthors}{Suzuki, et al.}


\begin{abstract}
This workshop brings together experts and practitioners from augmented reality (AR) and artificial intelligence (AI) to shape the future of AI-in-the-loop everyday AR experiences. With recent advancements in both AR hardware and AI capabilities, we envision that everyday AR---always-available and seamlessly integrated into users' daily environments---is becoming increasingly feasible. This workshop will explore how AI can drive such everyday AR experiences. We discuss a range of topics, including adaptive and context-aware AR, generative AR content creation, always-on AI assistants, AI-driven accessible design, and real-world-oriented AI agents. Our goal is to identify the opportunities and challenges in AI-enabled AR, focusing on creating novel AR experiences that seamlessly blend the digital and physical worlds. Through the workshop, we aim to foster collaboration, inspire future research, and build a community to advance the research field of AI-enhanced AR.
\end{abstract}




\maketitle

\section{Motivation}
%
Augmented Reality (AR) technologies are continuously advancing, both in terms of hardware and software.
Smaller form factors, extended battery life, and constant connectivity make it possible to results in a paradigm shift in how we think about AR: 
while current AR scenarios are focused on specialized applications such as productivity or maintenance \cite{gonzalez2024guidelines, gonzalez2016immersive}, we believe that \textbf{everyday AR} is increasingly feasible.
With everyday AR we refer to the making \textbf{AR always available to users}, enabling seamless interactions with the digital world, and potentially replacing or augmenting currently predominant technologies such as smartphone or desktop computing. 
Users will be able to ubiquitously query and interact with digital information, communicate with other users and virtual agents \cite{bovo2024embardiment}, and leverage AR for a majority of their interactions. 

This rise of everyday AR is fueled not just by improvements in hardware and software infrastructure, but is also heavily tied to recent advances in Artificial Intelligence (AI) and Machine Learning (ML).
Generative AI techniques~\cite{Goodfellow2014-sj}, for example, enable the creation of multi-modal digital content on-the-fly; and Large Language Models~\cite{Brown2020-ph} lead to advancements in various applications domains and make it feasible for users to interact with virtual agents through text and other inputs in a ``natural'' way.
Leveraging advances in those areas will enable researchers and practitioners to develop new and refine existing interactions for AR.
We believe that in order to make everyday AR feasible, we need to adopt an \textbf{AI-in-the-loop} approach, where the digital interactions and content continuously anticipate and adapt to users every-changing needs and context.
Thus, we believe the combination of AR and AI will enable us to move beyond monolithic applications in AR towards a truly user-centered and adaptive experience where both scene understanding and content generation becomes dynamic.

The goal of this workshop, which is an evolution of our previous \textit{XR and AI Workshop} at ACM UIST 2023~\cite{suzuki23uistworkshop}, is to bridge the fields of AR and AI and enable discussions around the feasibility and requirements for AI-enabled everyday AR.
Our previous workshop was broadly centered around the topic of what is possible when AR and AI are combined, the second iteration focuses on everyday AR interactions.
We plan to explore usage scenarios of this new paradigm, explore requirements and limitations of current and future AR with AI-in-the-loop, and identify and address common challenges of AR (e.g., accessibility, privacy) and AI (e.g., explainability, sustainability). 
In that sense, we consider this workshop as a natural evolution, not only we meet but also define a way forwards for the community. This workshop seeks to further refine the thinking of the community towards a shared vision, and should lay the foundation for future workshops at conferences, longer seminars, and work towards a reference paper to consolidate the research efforts. As such we propose a workshop very oriented to shared vision, and make a call for participation in which we ask participants to share how they imagine the future of AR as we move to a world where AR is to AI what screens have been to computers.

\subsection{Perspectives of Interest}
    This workshop welcomes HCI researchers and practitioners in AR, AI, machine learning, and computational interaction domains to share diverse perspectives and expertise. 
    There are several domains that are not fully explored yet in the literature of XR and AI~\cite{hirzle2023xr}. We plan to discuss the topics that include but are not limited to the following areas: 
    
    \subsubsection*{\textbf{Adaptive and Context-Aware AR}}
    Unlike traditional user interfaces on smartphones or computers, AR can embed virtual elements into the user's physical world. However, without careful design, such interfaces can become overwhelming and distract users unnecessarily. Context-aware AR aims to seamlessly blend these virtual elements by automatically adapting them based on users' needs, context, and environment~\cite{lindlbauer2022future, lindlbauer2019context, grubert2017pervasive}. Recent research has explored various aspects of context-aware AR, such as using computational optimization methods to blend AR interfaces into physical objects~\cite{han2023blendmr}, leveraging everyday objects to provide affordances for virtual interactions~\cite{he2023ubi, jain2023ubi}, and applying adaptive AR for specific use cases like improving accessibility for people with low vision~\cite{lee2024cookar}. We are interested in exploring how to expand current research into broader everyday AR applications by leveraging advanced AI capabilities. These capabilities can help AR systems better understand various types of contextual information, such as room geometry, the affordances of physical objects, and user activities, enabling more seamless and adaptive experiences.
    
    \subsubsection*{\textbf{Always-on AI Assistant by Integrating LLMs and AR}}
    Large Language Models (LLMs) have significant potential to augment AR experiences. Existing LLM interfaces, like ChatGPT, require explicit user interaction, but integrating LLMs into everyday AR interfaces allows for always-on AI assistants that can implicitly support user activities and integrate naturally into the environment. For example, rather than requiring users to type questions on a screen, the integration of LLMs and AR can facilitate seamless interactions with digital content, making AR more intuitive. We are interested in exploring how the integration of LLMs can extend beyond simple screen-based interactions to enable new, multimodal applications within AR environments. Recent work has demonstrated how augmented object intelligence can make the physical world interactable in AR~\cite{dogan2024augmented}, on-demand mixed reality document enhancement~\cite{gunturu2024realitysummary}, gaze and gesture-based question answering~\cite{lee2024gazepointar}, and everyday assistance based on current activities through wearable devices~\cite{arakawa2024prism}. We are interested in exploring how this integration can be advanced with the multimodal capabilities of LLMs. Moreover, we believe there are significant opportunities beyond just textual output modalities, utilizing embedded visual, tangible, and spatial modalities that are unique to AR interfaces.
    
    \subsubsection*{\textbf{AI-Assisted Task Guidance in AR}}
    AI-assisted task guidance offers significant advantages over traditional video-based instruction \cite{castelo2023argus}. While videos provide static, one-way demonstrations, AI systems can analyze a user's movements and activities in real-time, offering personalized feedback and corrections. We are interested in exploring how this interactive approach can allow learners to receive immediate, tailored guidance on their form, timing, and technique and what is the potential to accelerate learning, reduce the risk of developing bad habits, and enhance overall performance. Beyond full body activities such as physical fitness or sports training, we are also interested in manual task assistance such as cooking, repair or assembly. Additionally,we are interested in understanding how AI-based systems can adapt to a user's progress, gradually increasing task complexity and providing encouragement, creating a more engaging and effective learning experience than video watching alone.
    
    \subsubsection*{\textbf{AI-in-the-Loop On-Demand AR Content Creation}}
    Generative AR has the potential to transform the creation of interactive AR content. With recent advances in generative AI, it is now possible to generate on-demand AR content in real time. By leveraging advances in image generation, such as GANs and Stable Diffusion, generative AI can facilitate content creation, 3D and avatar animation~\cite{ahuja2021coolmoves}, and immersive environment design~\cite{gal2014flare}. Beyond static image generation, generative AI can also be used to create interactive AR content by leveraging code generation capabilities of LLMs~\cite{de2024llmr, aghel2024people}. Furthermore, by understanding real-world objects and content, users can make static content interactive in AR~\cite{chulpongsatorn2023augmented}. We are particularly interested in exploring how generative AI can leverage the unique physical and spatial aspects of AR, such as 3D scene understanding and spatial interactions, for various kinds of generated content, including text, visuals, motion, video, and 3D objects, allowing for a richer and more dynamic AR experience. We want to discuss the challenges and possibilities of integrating these capabilities into AR environments.
    
    \subsubsection*{\textbf{AI for Accessible AR Design}}
    AI presents transformative potential for enhancing accessibility in AR design and experiences, addressing the exclusion faced by individuals with hand-motor impairments. Currently, these users are largely shut out from both creating and engaging with AR content due to the prevalent use of traditional input methods. Tools like Unity, essential for XR design, demand bimanual keyboard-mouse input, while XR experiences typically require handheld controllers or precise hand movements—both significant barriers for those with motor disabilities. To overcome these challenges, we propose exploring AI-driven alternative input modalities such as voice commands coupled with gaze tracking, facial expression recognition \cite{taheri2021exploratory}, and AI-interpreted subtle body movements. Furthermore, AI can enhance the XR experience by adapting content difficulty, adjusting interfaces, and modifying narratives to match individual capabilities. This approach aims to create inclusive solutions that not only make XR design and experiences accessible to a broader audience but also innovate interaction paradigms for all users, potentially revolutionizing how we create and interact with virtual worlds. 
    
    \subsubsection*{\textbf{Real-World-Oriented AI Agents}}
    We are also interested in exploring how we can design AI agents that enhance our physical world. The future of AI agents should bridge the gap between the digital and physical worlds by understanding spatial relationships, object affordances, and user activities. For example, such AI agents could assist users in navigating their environment, supporting complex everyday tasks, or providing personalized assistance. Recent work has shown promising developments in creating such agents that guide and visualize users' attention through mixed reality avatars~\cite{thanyadit2023tutor}. We aim to discuss the possibilities of designing AI agents for AR that are aware of and responsive to the real-world context.

\section{Organizers}

\textbf{Ryo Suzuki} is an Assistant Professor in the ATLAS Institute and Department of Computer Science at the University of Colorado Boulder. His research interests focus on augmented reality and tangible user interfaces. His research aims to transform everyday environments into dynamic physical and spatial media with the power of AI and AR.
His website is \url{https://ryosuzuki.org/}.

\vspace{0.2cm}\noindent
\textbf{Mar Gonzalez-Franco} is a Research Manager at Google, where she leads Blended Interaction Research and Devices (BIRD) group. She has pioneering research in the field of extended reality in various domains such as human perception of virtual avatar, haptic interfaces for VR, multi-device interactions for XR, and ML-enabled novel experiences. Her website is \url{https://margonzalezfranco.github.io/}.

\vspace{0.2cm}\noindent
\textbf{Misha Sra} is an Assistant Professor in the Department of Computer Science at the University of California, Santa Barbara, where she leads the Human-AI Integration Lab (HAL). Her research focuses on the design of Human-AI systems in XR to augment human potential which includes understanding human psychology, building specialized hardware for data collection, designing new AI models, creating interactive human-AI systems, and conducting large scale evaluations of novel applications. Her website is \url{https://sites.cs.ucsb.edu/~sra/}.

\vspace{0.2cm}\noindent
\textbf{David Lindlbauer} is an Assistant Professor at the Human-Computer Interaction Institute at Carnegie Mellon University where he leads the Augmented Perception Lab. His research focuses on creating and studying enabling technologies and computational approaches for adaptive user interfaces to increase the usability of AR and VR interfaces, with applications in casual interaction, productivity, health and robotics. His website is \url{https://davidlindlbauer.com/}

\section{Workshop Format and Plans}

\subsection{In-Person Workshop Structure}
The workshop will be a one-day, in-person event featuring a variety of activities, including small-group discussions, paper presentations, keynote talks from invited experts, and hands-on design activities to explore the future of AI-enabled everyday AR. One of the primary goals is to foster a strong community around this emerging topic, leveraging the networking opportunities provided by face-to-face interactions, such as a group lunch and dinner. The design sprint activities will particularly benefit from the collaborative nature of in-person participation, allowing teams to work flexibly with shared digital and physical materials for prototyping. To ensure an intimate and engaging workshop experience, we plan to cap attendance at 30 participants. This small group size will help create an environment that encourages meaningful interactions and effective collaboration.

\subsection{Workshop Activities}
\begin{figure*}[t]
    \centering
    \includegraphics[width=\linewidth]{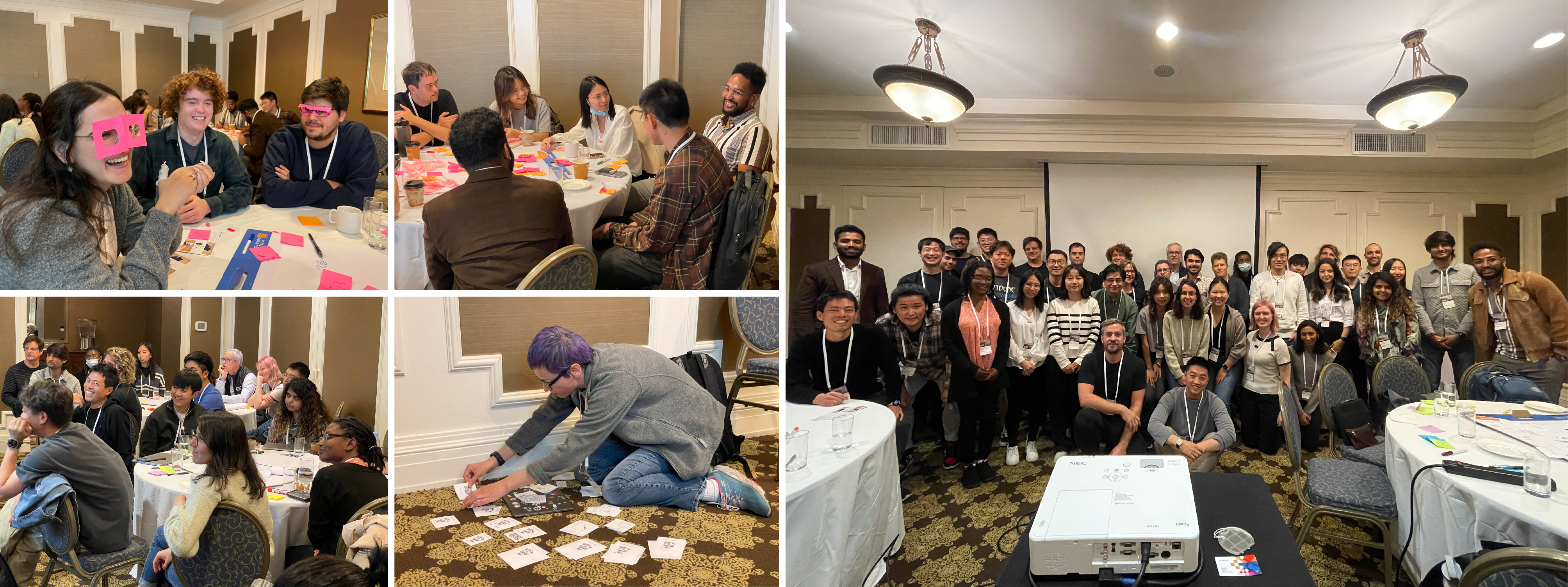}
    \caption{Photos of the XR+AI workshop at ACM UIST 2023. Around 50 participants engaged in a variety of activities such as brainstorming and scenario acting, paired with keynotes, lightning talks and panel discussions.}
    \Description{A collage of photos from the XR+AI workshop at ACM UIST 2023. Participants are engaged in activities such as brainstorming, wearing playful design prototypes, arranging cards on the floor, and posing for a group photo in a large meeting room. The group is diverse, with attendees participating in both seated and active group tasks.}
    \label{fig:uist_workshop}
\end{figure*}

The workshop will include a series of interactive and talk-based activities, similar to our successful workshop at ACM UIST 2023, as shown in Figure~\ref{fig:uist_workshop}. In the previous UIST workshop, we had a total of 40 participants from both academia (32) and industry (8), representing diverse countries, genders, fields, backgrounds, and levels of experience. They actively engaged in lightning talks, prototyping sessions, discussions, and design activities. For the upcoming iteration, we will retain these participant-centered activities while also expanding the workshop to include position paper presentations from attendees and a keynote talk to inspire a vision for the future of AI and AR in everyday contexts. Below, we outline our current workshop format and plans.

\subsubsection*{\textbf{Introductions and Lightning Talks}}
The organizers will kick off the workshop with introductions and each participant's lighting talks highlighting their related research experience. To start community building, we will facilitate pre-workshop engagement to cultivate and propose an initial set of themes through the creation of a shared slide deck that highlights each participant's background and interests.
We will invite two keynote speakers who will deliver presentations to provoke further discussions. 

\subsubsection*{\textbf{Position Paper Presentations}}
In this workshop, we will announce a public call for position papers from the CHI community, inviting researchers and practitioners to contribute their insights. The purpose of these talks is to provide an opportunity for attendees to share their opinions and positions on their recent and ongoing work related to AR and AI. We aim to foster a collaborative environment where participants can present radical ideas, exchange feedback, and inspire new directions for research. For each accepted position paper, we plan to allocate 15-20 minutes for presentations. Additionally, with the consent of the participants, we plan to publish these position papers after the workshop to make the insights available to a broader audience and promote further discussion.

\subsubsection*{\textbf{Keynote Talks}}
We will also invite distinguished speakers to present keynote talks, with the goal of sharing forward-thinking visions and ideas with the audience. These talks are intended to inspire and set future directions for the field, similar to the UIST Vision Talks. Each keynote session will be approximately one hour long, providing ample time for in-depth exploration of ideas. We also plan to publish the title, speaker bio, and abstract on the workshop website to provide attendees with context and background information ahead of the event.

\subsubsection*{\textbf{Participant Activities}}
Our workshop aims to provide a balanced mix of presentations and interactive participant activities to foster engagement and collaboration among attendees. To achieve this, we have organized several activities designed to encourage hands-on exploration and creative problem-solving. These include: \textit{1) Hackathon and Rapid Prototyping:} Teams use low-fidelity materials (paper, cardboard, markers) to mock up an XR interface or device. Teams incorporate how AI might enhance the user experience. Teams present prototypes in a "science fair" style showcase. \textit{2) Design Sprint:} We will challenge participants to redesign a common XR experience for users with specific disabilities. Users may integrate AI to propose adaptive interfaces or alternative input methods. \textit{3) XR-AI Mind Mapping:} We will create a large, shared mind map of potential XR+AI applications across industries and encourage participants to add ideas throughout the day. 

\subsubsection*{\textbf{Theme Organization and Discussion}} 
Participants will collectively extrapolate themes from the presentations and shared readings. The discussed themes will be combined with the initial set of topics developed prior to the workshop to serve as guiding directions for the first round of discussions. Once a preliminary list of discussion topics has been defined, each topic will be assigned a ‘table.’ During the session, participants will rotate between tables to engage in focused discussions of topics of their choice. One participant at each table will be designated as the discussion mediator, whose responsibilities will involve guiding and documenting the discussion. The second session on theme organization will begin with lightning talks by the discussion mediators summarizing earlier conversations. Participants will then collectively revisit the discussion topics, reorganizing accordingly based on the results of the first session and expert perspectives. With the refined list of topics, the remainder of the session will follow the same format as the first round of discussions. 

\subsubsection*{\textbf{Defining Future Challenges and Research Directions}} 
In this final discussion session, we will begin by regrouping and refining the list of discussion topics based on the results of the discussion sessions. The final workshop session will focus on summarizing the workshop findings and defining next steps. First, the organizers will provide a recap of the workshop activities, including the defined themes from the pre-workshop activities and a summary of the morning and afternoon discussion outcomes. The floor will then be opened for participants to contribute their reflections on the workshop discussion. A final discussion will be held around potential future directions, such as follow-up workshops and publications.

\section{Call for Participation}
This one-day, in-person CHI 2025 workshop invites researchers, practitioners, and students interested in exploring the future of AI-enabled augmented reality (AR) to join us in shaping a vision for "Everyday AR." With recent advancements in AI and hardware technologies, the idea of always-available, seamlessly integrated AR is becoming increasingly feasible. Our workshop will focus on how AI can enable adaptive, user-centered AR experiences that seamlessly integrate into everyday life. Participants in the workshop will discuss with a variety of topics, including adaptive and context-aware AR, generative AR content creation, LLM-powered always-on AI assistants, AI-driven accessibility, and the development of real-world-oriented AI agents. Through a combination of keynote talks, participant-led presentations, and hands-on activities, attendees will have the opportunity to co-create visions of the future, exchange ideas, and contribute to the development of truly adaptive and AI-enhanced AR. Participants will also learn about the state-of-the-art in this space, hear perspectives from experts in AR and AI, and participate in collaborative prototyping and design activities. We aim to create a shared vocabulary and research agenda for AI-driven AR, inspiring new directions and fostering future collaborations. To apply for participation, interested individuals should complete an application form indicating their background, experience, and interests in AR and AI by February 15th. Applicants are encouraged to optionally submit a position paper of up to 2 pages (without references) in the ACM double-column format. Participants will be selected based on their application materials, with a focus on the relevance of their research or disciplinary background to the workshop topic, and a goal of creating a diverse and academically engaging environment. 

\subsection{Registration Requirements}
Note that per CHI 2025 policies, workshop participants (including at least one author of each position paper, for applications that optionally include position papers) must attend the workshop in Yokohama, and all participants must register for both the workshop and for at least one day of the conference.


\subsection{Website and Participant Recruitment} 
The website of the workshop can be found at \url{https://xr-and-ai.github.io/}.
We will distribute a Call for Participation on the website through our research groups, professional networks, social media platforms, previous attendees of the UIST workshop, and HCI-related mailing lists.
Prospective attendees will be invited to complete a short form describing their background, experience, and interests. 
Workshop organizers and invited external experts will collectively curate an inclusive selection of participating researchers based on the diversity of career levels, research interests, world locations, and genders, while endeavoring to match expertise in experience in XR and AI research.

\subsection{Plans to Publish Workshop Proceedings} 
We plan to publish the content submitted by participants, including position papers, as workshop proceedings to ensure accessibility and foster ongoing collaboration. These materials will be hosted on our website, and we will encourage participants to also publish their work on open-access platforms such as CEUR-WS or arXiv.

\balance

\bibliographystyle{ACM-Reference-Format}
\bibliography{references}

\end{document}